\documentclass[twocolumn]{aastex6}
\usepackage{graphicx}
\usepackage{grffile}
\usepackage{bm}
\usepackage{psfrag}
\usepackage{mathrsfs}
\usepackage{placeins}
\usepackage{amssymb,amsmath}
\newcommand{\be}{\begin{eqnarray} }
\newcommand{\ee}{ \end{eqnarray} }

\renewcommand{\mnras}{MNRAS}
\renewcommand{\araa}{ARAA}
\renewcommand{\pasp}{PASP}

\renewcommand{\nat}{Nature}
\renewcommand{\be}{\begin{align}}
\renewcommand{\ee}{\end{align}}

\begin{document}
\title{Radio-interferometric monitoring of FRB~131104:\\   A coincident AGN flare,  but no evidence for a cosmic fireball }
\shorttitle{Fast radio burst counterparts}

\shortauthors{R.~M. Shannon \& V.~Ravi}
\author{R. M. Shannon\altaffilmark{1,2,4}}
\email{ryan.shannon@csiro.au,vikram@caltech.edu}
\author{V. Ravi\altaffilmark{3,4}}
\altaffiltext{1}{CSIRO Astronomy and Space Science, Australia Telescope National Facility, Box 76, Epping NSW 1710, Australia}
\altaffiltext{2}{International Centre for Radio Astronomy Research, Curtin University, Bentley WA 6102,  Australia}
\altaffiltext{3}{Cahill Center for Astronomy and Astrophysics, MC 249-17, California Institute of Technology, Pasadena, CA 91125, USA}
\altaffiltext{4}{The authors contributed equally to this work}




\begin{abstract}
The localization of fast radio bursts (FRBs) has been hindered by the poor angular resolution of the detection observations and inconclusive identification of transient or variable counterparts.
Recently a $\gamma$-ray pulse of $380$~s duration has been associated with the fast radio burst FRB~131104.
  We report on radio-continuum imaging observations of the original localization region of the FRB, beginning three days 
after the event and comprising 25 epochs over 2.5~yr. Besides probabilistic arguments that suggest that the association between the $\gamma$-ray transient and the FRB 
is not compelling, we provide upper limits on a putative radio afterglow of this transient that are at odds with standard models for its progenitor.
We further report the discovery of an unusual variable radio source spatially and temporally  coincident with FRB~131104, but not spatially coincident with the 
$\gamma$-ray event. 
The radio variable flares by a factor of $3$  above its long term average within $10$~d of the FRB at 7.5\,GHz, with a factor-of-$2$ increase at $5.5$~GHz. 
Within our observations, the variable has persisted with only modest modulation and  never approached the flux density observed in the days after the FRB. 
We identify an optical counterpart to the variable. Optical and infrared photometry, and deep optical spectroscopy, suggest that the object is a narrow-line radio AGN. 

\end{abstract}

\keywords{fast radio bursts --- radio continuum: general --- gamma rays: general --- galaxies: active --- black hole physics }

\section{Introduction}

Fast radio bursts (FRBs) represent a new phenomenological class of astrophysical transient. 
They are bright  ($\gtrsim$Jy peak flux density) pulses of radio emission that show  the effects of propagating though large column densities of plasma:  dispersion through ionized plasma, multipath propagation due to inhomogeneities in the plasma  \cite[][]{2007Sci...318..777L,2013Sci...341...53T}, and Faraday rotation due to magnetization of the plasma \cite[][]{2015Natur.528..523M}. 
The column densities exceed predictions for the Galaxy, suggesting that the FRBs are extragalactic and possibly cosmological in origin \cite[][]{frb150807paper}.
They have hitherto only been detected using single-dish telescopes, which have poor angular resolution.
Only one FRB has been found to repeat \cite[][]{2016Natur.531..202S}, greatly enhancing prospects for its localization in follow-up observations. 
For the remaning FRBs, which have not repeated, attempts at localization have relied on detecting counterpart multi-wavelength transients that might be expected if the FRBs arise from cataclysmic explosions or outbursts.
A claimed association of FRB~150418  with a centimeter-wavelength afterglow and  host  galaxy \cite[][]{2016Natur.530..453K} has been disputed and instead attributed to common AGN variability, either intrinsic \cite[][]{2016ApJ...821L..22W,2016ApJ...824L...9V}, or caused by Milky Way scintillation 
\citep{2016ApJ...824L...3A,2016arXiv161009043J}. 
With so little detail on the locations of FRBs, theories for their production and 
sources are understandably varied, ranging from ultabright pulses from pulsars \citep{2016MNRAS.457..232C}, to  cosmic strings \citep{2012PhRvD..86d3521C}.

We detected FRB~131104 \cite[][]{2015ApJ...799L...5R} with the 64-m Parkes radio telescope in the direction of the Carina dwarf spheroidal galaxy  (Car dSph), $100$~kpc distant from Earth. 
The FRB has an electron column density, measured in units of dispersion measure (DM), of $779.0\pm0.2$ pc~cm$^{-3}$ and shows evidence for temporal broadening associated with multipath propagation.  
Despite its detection in a targeted observation of the Car dSph, we have no evidence to associate the FRB with that galaxy.
The FRB has not repeated in $\approx 100$~hr of follow-up observations at Parkes. 

A $\gamma$-ray transient, Swift J0644.5$-$5111, has recently been associated with the FRB at the $3.2\sigma$ to $3.4\sigma$ confidence level 
\cite[][]{carinaswift}. The emission was detected in an off-axis position with the {\em Swift} satellite's Burst Alert Telescope \cite[BAT;][]{2005SSRv..120..143B}, coincident with the FRB in position and time. The transient duration was  $\sim380$~s, with an inferred energy output 
of $5\times10^{51}$\,erg. \citet{carinaswift} suggested that the $\gamma$-ray emission (assumed to be associated with 
this FRB) was generated by shocked relativistic plasma in a cosmological explosion, or in an accretion episode associated with a supermassive 
black hole. We discuss the claimed association between Swift J0644.5$-$5111 and FRB~131104 in Section \ref{sec:swift}, 
addressing specifically the mismatch (noted by DeLaunay et al.) between the low rate of such transients observed by {\em Swift}/BAT and the 
high FRB all-sky rate.

Here we report on a centimeter-wavelength radio monitoring campaign of the Parkes localization region of FRB~131104, 
and the discovery of an unusual,  variable radio source (AT~J0642.9$-$5118) that flares coincident in time and location with FRB~131104. 
AT~J0642.9$-$5118 is not coincident with  Swift J0644.5$-$5111; indeed, our observations exclude any bright radio afterglow of Swift J0644.5$-$5111. 
    In Section \ref{sec:radioobs}, we present radio observations of the field and the light curve of AT~J0642.9$-$5118. 
    In Section \ref{sec:opticalobs}, we present an optical characterization of AT~J0642.9$-$5118.
We discuss the implications of our observations  
in Section \ref{sec:discuss},  and conclude the paper in Section \ref{sec:conclude}.

\section{The $\gamma$-ray transient coincident with FRB~131104}
\label{sec:swift}

Swift J0644.5$-$5111 was discovered within the 15\arcmin~diameter half-power circle of the beam (\#5) of the Parkes 21-cm multibeam 
receiver in which FRB~131104 
was detected, 6.2\arcmin~from the beam center. Its position at the edge of the BAT field of view, illuminating only 2.9\% of detectors, 
resulted in a $4.2\sigma$ detection in the image plane despite its high fluence of $4\times10^{-6}$\,erg\,cm$^{-2}$. 
Assuming a  distance of 3.2\,Gpc for FRB~131104 based on comparing its extragalactic DM with models for the ionized content of the Universe, 
the isotropic energy output of Swift J0644.5$-$5111 was $5\times10^{51}$\,erg, with a duration of $\sim380$\,s. This is somewhat longer, 
and somewhat less energetic than most long-duration gamma-ray bursts (GRBs) detected by {\em Swift} \cite[][]{2009ARAA..47..567G}, 
but is inconsistent with other GRB subtypes (e.g., ultra-long GRBs). \citet{2016arXiv161103848M} consider it likely 
that, largely independent of the source model, a radio afterglow would have been present. We constrain such an afterglow in Section~5.2 using our observations.

\citet{carinaswift} estimate a significance for the association between FRB~131104 and Swift J0644.5$-$5111 of between $3.2\sigma$ and $3.4\sigma$, based on the estimated false positive rate in a large collation of BAT archival data. This corresponds to an odds ratio of between  $\approx$ 600:1 
and 1800:1. Following an argument made by \cite{2016ApJ...824L...9V}, we compare this odds ratio with the expected number of FRBs that 
exhibit similar counterparts, which we can estimate by comparing the detection rate of events such as Swift J0644.5$-$5111 with the FRB rate. 

There is an inconsistency between the inferred all-sky $\gamma$-ray pulse rate and the FRB rate, as noted by  \citet{carinaswift}, that also calls into question the association.
If Swift J0644.5$-$5111 had occurred in the region of the BAT field of view with $>90\%$ coding, it would have resulted in an image-based burst 
trigger.
 \citet{carinaswift} estimate that the rate of long-duration image-triggered events, presumably similar to Swift J0644.5$-$5111, is 25\,yr$^{-1}$.  
We make the conservative assumption that these events all have FRB counterparts, regardless of their fluence or classification.
The 100\% 
coding region of BAT is $\approx1000$\,deg$^{2}$ \cite[][]{2005SSRv..120..143B}, which we (conservatively) equate with the $>90\%$ coding 
region.
 In this region,  we predict that BAT should have been 
sensitive to the counterparts of between $8800$ and $17600$ FRBs in a year.  
We calculate this using the  fluence-complete FRB rate of 2500\,sky$^{-1}$\,day$^{-1}$ events 
with fluences $>2$\,Jy\,ms \cite[][]{2015MNRAS.447.2852K}, 
and assume  both that the FRB source counts are consistent with a Euclidean universe,  and that Swift obtains a $>50\%$ observing duty cycle.
Thus, the odds ratio of \citet{carinaswift} observing their  counterpart is the ratio of the $\gamma$-ray event rate to the radio event rate. 
This places the odds ratio at between approximately 350:1 and 700:1. 

Therefore, the odds ratio of   FRB~131104 having a $\gamma$-ray counterpart (based on the disparity of the $\gamma$-ray pulse and FRB rates), 
and the odds ratio  of Swift J0644.5$-$5111 being 
associated with FRB~131104 \citep[the calculation presented in][]{carinaswift},  are comparable. 
This demonstrates that a true association is not significantly more likely than the probability of an unassociated occurrence. 
This issue was qualitatively acknowledged by \citet{carinaswift}.
To reconcile the event rates would require FRB~131104 to be of a fundamentally different, much rarer class than the other FRBs.  

 Furthermore, the false alarm probabilities of such unassociated occurrences 
given by \citet{carinaswift} are likely underestimated. The calculations relied on estimating the background rate of $4.2\sigma$ image-plane 
detections, when lower-significance detections may still have exceeded their final false-alarm probability threshold of $3\sigma$. Their background 
rate was also only calculated for events with 200\,s to 400\,s durations, whereas they may still have claimed a counterpart discovery with either 
a shorter or longer event coincident with FRB~131104. The false alarm rate for all the possible associations that \citet{carinaswift} could 
have claimed is hence likely higher than was estimated. 

\section{Radio-interferometric Observations}
\label{sec:radioobs}

\begin{figure}[!ht]
\begin{center}
\begin{tabular}{cc}
\includegraphics[scale=0.4,trim={0 0.7in 0 0},clip]{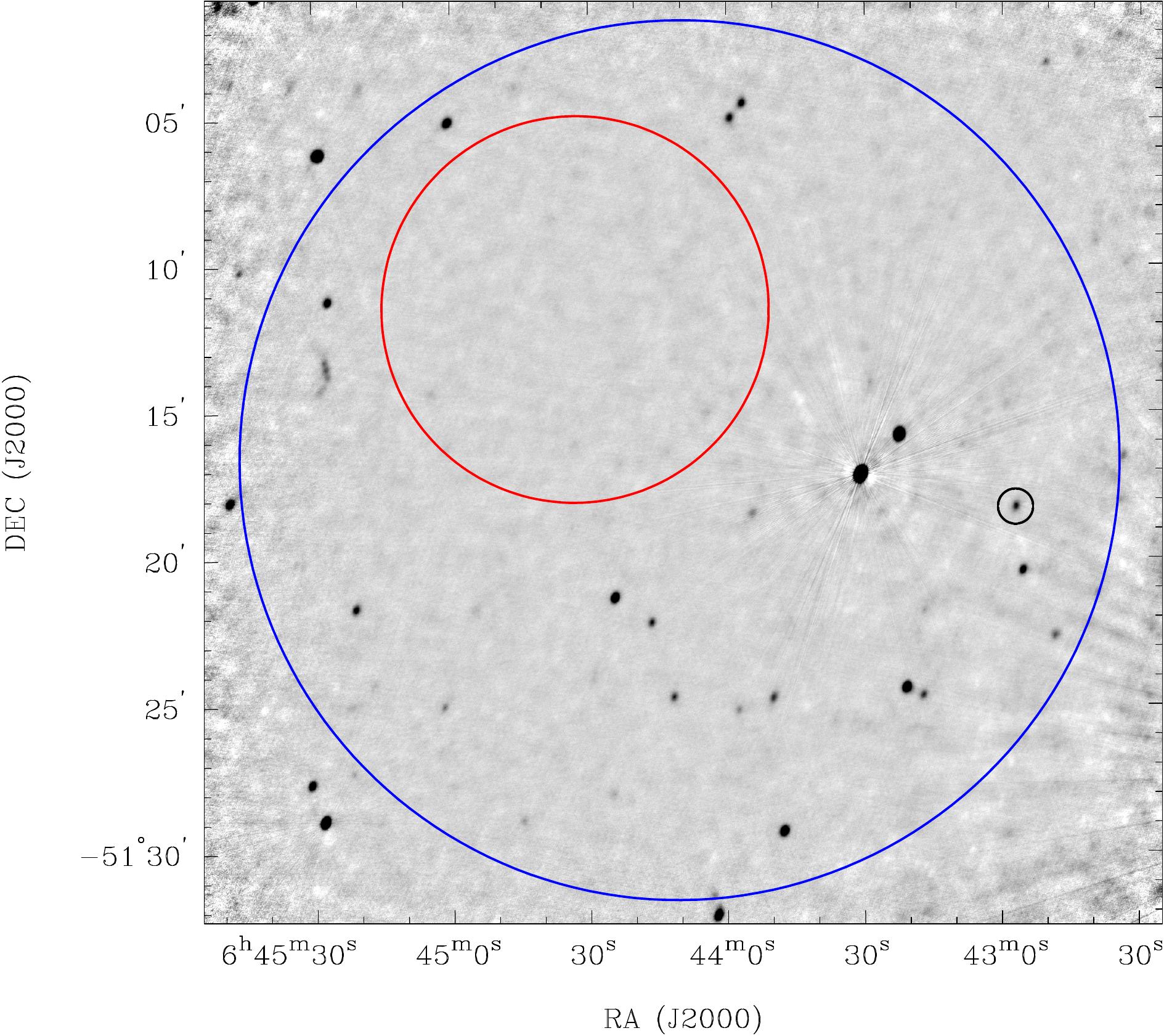}    \\ \includegraphics[scale=0.4,trim={0 0 0 0.05 n},clip]{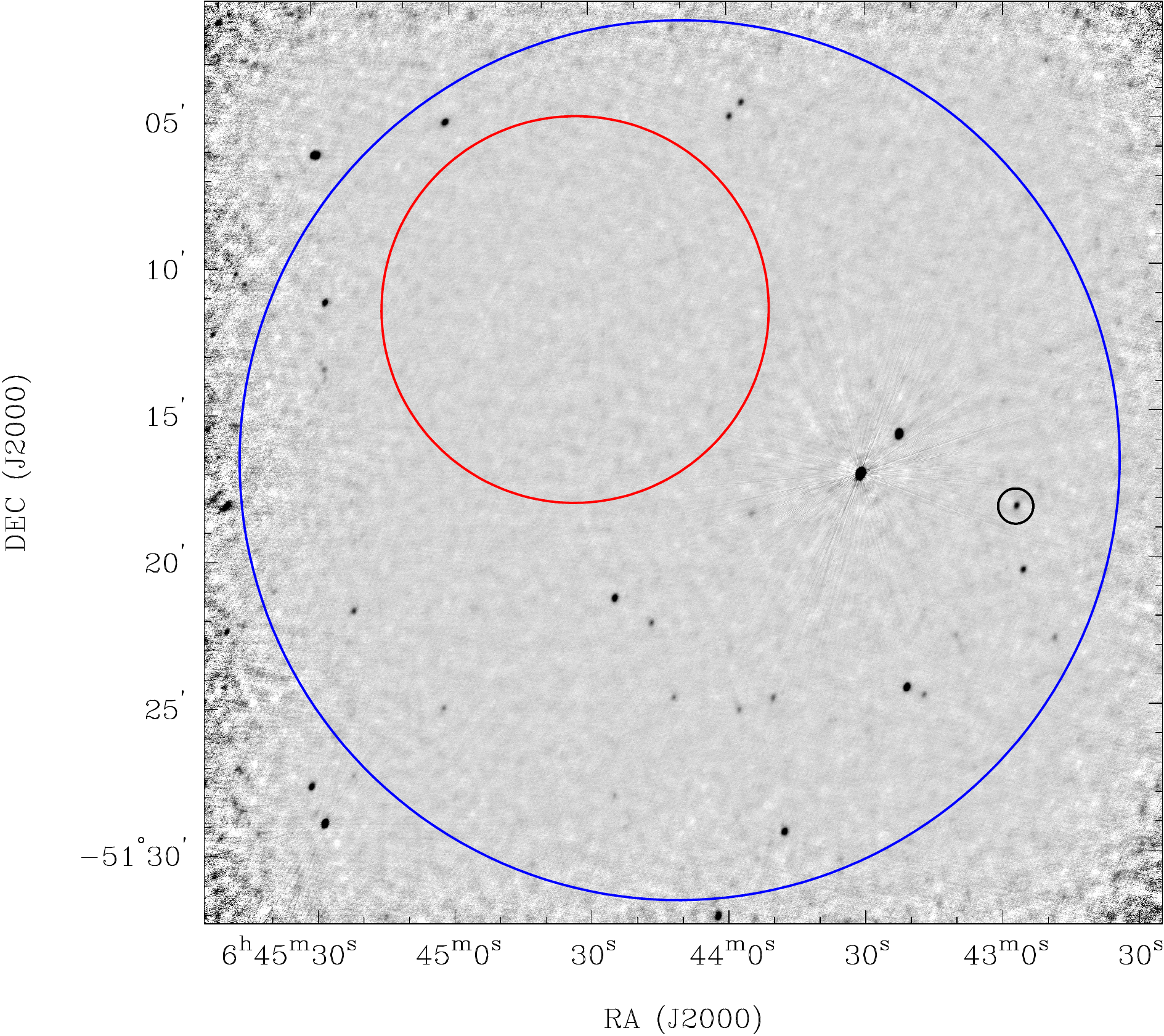}\\
\end{tabular}
\end{center}
\caption{ 
 \label{fig:mosaic}  
Radio-continuum images of the field surrounding FRB~131104 in the 5.5 GHz band (top) and 7.5 GHz band (bottom).  The blue circle shows the beam of  Parkes telescope (to twice the half power point, which is approximately the first null in the beam pattern). The 5.5~GHz and 
7.5~GHz image rms flux densities within the blue circle are $15~\mu$Jy beam$^{-1}$ and $20~\mu$Jy beam$^{-1}$ respectively.
The red circle shows the 90\% confidence region for the Swift transient.    The black circle shows  the position of the unusual variable AT~J0642.9$-$5118.  In both plots, the grayscale ranges linearly from $-100$ to $500~\mu$Jy~beam$^{-1}$. }
\end{figure}

We commenced monitoring the field of FRB~131104 with the Australia Telescope Compact Array (ATCA) $3$~d after the FRB was detected at Parkes. 
Our observations were conducted over 25 epochs spanning 2.5\,yr. 
Visibilities were computed using the Compact Array Broadband Backend  \cite[][]{2011MNRAS.416..832W} over two $2$~GHz width tuneable bands, centered at $5.5$ and $7.5$~GHz.
A $42$-pointing mosaic was necessary to cover to twice the half-power beam point of Parkes observations (which is the first null in the primary beam)  at the highest frequency of the ATCA observations. 
This was especially crucial because of the possibility of a population of ultrabright FRBs that could be detected in the outer main beam or sidelobe of the telescope \cite[][]{2016ApJ...830...75V}.

Observations were conducted in a variety of array configurations, with maximum baseline lengths varying between $214$~m and $6$~km.
Usually $6$ antennas were available, but some observations were conducted with $5$ antennas (particularly in the lower resolution arrays where inclusion of a sixth very distant antenna complicates imaging), and one with $4$. 
The lower spatial resolution observations suffered from higher noise, but other problems such as source confusion were not a problem because the field is relatively sparse.
Data were bandpass calibrated using  observations of either PKS~0823$-$500 or PKS~1934$-$638, and flux calibrated using the latter.
 Phase calibration was conducted with regular observations of the unresolved radio galaxy J0625$-$5438.
 Data were reduced using the {\tt miriad} data reduction package \cite[][]{1995ASPC...77..433S}.
 The visibilities for each pointing were imaged and deconvolved independently (using multi-frequency synthesis and cleaning)  and then combined to form a composite image.  
Noise levels were typically $30~\mu$Jy beam$^{-1}$ in the mosaicked observations. 
 We investigated the role of self calibration (both phase-only and amplitude-and-phase self calibration) on our flux-density measurements. We found that while self calibration improved image fidelity it did not significantly alter flux-density measurements\footnote{\cite{2016arXiv161009043J} noted a $\approx 10\%$ downward bias in flux-density measurements in mosaicked data sets.  We found that this was mitigated by imaging with the source of interest at the reference pixel of the image.}.
  
Figure \ref{fig:mosaic} shows composite images  formed from the $5.5$~GHz (top)  and  $7.5$~GHz (bottom) observations of the field.
The rms noise levels in the two images are, respectively, $14$ and $19~\mu$Jy beam$^{-1}$. 
The width of the primary beam of Parkes, to the first null, is shown as the blue circle. 
The 90\% containment region for Swift J0644.5$-$5111 is shown as the red circle. There are no sources within this region in either the 
mosaics of all our data shown in Figure \ref{fig:mosaic} or in individual epochs, 
allowing us to place $5\sigma$ limits on persistent sources at $5.5$~GHz and $7.5$~GHz of $70$ and $100~\mu$Jy respectively.

\begin{figure}[!ht]
\begin{center}
\begin{tabular}{c}
\includegraphics[scale=0.57,angle=-90]{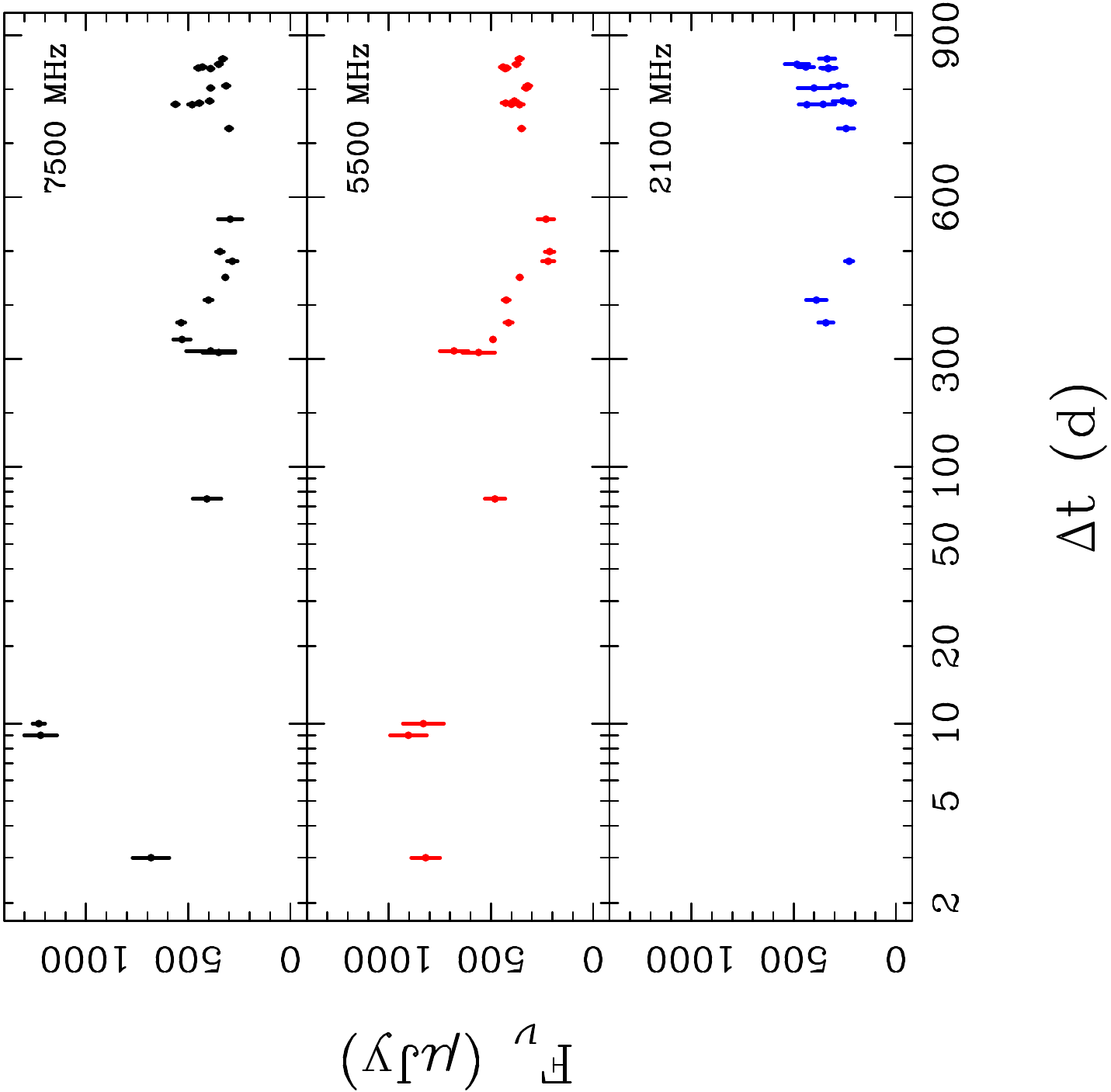}\\
\end{tabular}
\end{center}
\caption{ 
 \label{fig:light_curve}  
Radio light curve for the variable radio source AT~J0642.9$-$5118.  From  the lower to upper panels we show the flux density measured 
in the bands centered at $2.1$~GHz, $5.5$~GHz, and $7.5$~GHz. The x-axis is the time since FRB~131104. 
For $\Delta t < 100$~d we show the light curve on a logarithmic time axis.  For $\Delta t  > 100$~d the axis is linear. 
}
\end{figure}

Within the field of view, we have identified a strongly variable source, which we refer to as AT~J0642.9$-$5118. 
The location of the source on the sky is (J2000) $\alpha = 6^{\rm h}42^{\rm m}57^{\rm s}.154(3),  \delta=-51^\circ18^{'}17^{"}.70(7)$.
The light curve for the source is presented in Figure \ref{fig:light_curve}.  
In the week after the occurrence of FRB~131104, the source brightens by a factor of $2$, exceeding $1200~\mu$Jy in the 7.5~GHz band. 
During the brightening, the spectrum also inverts. Other sources in the field do not show this level of variability, suggesting that 
mis-calibration has not introduced the flux variation. 

After identifying AT~J0642.9$-$5118, we conducted more sensitive single-pointing observations at $2.1$, $5.5$, and $7.5$~GHz.  
Observation and data reduction in the 2.1~GHz band followed the same procedures as in the mosaicked observations.  In the $2.1$~GHz band, 
the major differences were that only $2$~GHz of bandwidth was available, and phase calibration used the radio galaxy PKS~0647$-$475. 
For these targeted observations, image rms noise was typically $30~\mu$Jy~beam$^{-1}$ in the $2.1$~GHz band, and $10~\mu$Jy beam$^{-1}$ in the higher-frequency 
images. 

There was a modest re-brightening of AT~J0642.9$-$5118 approximately $300$~d after the initial flare. 
After the initial flare the flux density at $7.5$~GHz has a mean value of $395~\mu$Jy and an rms value of $80~\mu$Jy, suggesting that 
the $1.2$~mJy event is a $10\sigma$ event temporally coincident with the FRB;  at $5.5$~GHz the mean flux density has 
been $390~\mu$Jy with an rms value  $100~\mu$Jy after the flare. 
The modulation index in the $5.5$~GHz ($7.5$~GHz) band is $0.3$  ($0.2$) when excluding the first three observations and $0.4$ ($0.5$)  when including them.

\section{Optical observations of AT~J0642.9$-$5118} \label{sec:opticalobs}

\subsection{Imaging}

Following the identification of  AT~J0642.9$-$5118, we obtained images of its optical counterpart with the Gemini Multi-Object Spectrograph 
\citep[GMOS;][]{gmos} at the Gemini-South telescope. Our observations were conducted on MJD~57141 in the $g'$ and $r'$ bands using the 
Hamamatsu CCDs \citep{gmosupgrade}, under photometric conditions with 0.6\arcsec~FWHM seeing. Four dithered exposures were taken in 
each band, totalling 2617\,s in the $g'$-band and 2322\,s in the $r'$-band. We reduced the data using the standard GMOS pipeline tasks in the Gemini IRAF 
package. We used facility bias and flat-field exposures nearest in time to our observations to correct the data, and co-added all images following 
subtraction of dithering offsets. Astrometric corrections were applied to the images using D. Perley's {\it autoastrometry} software\footnote{\url{http://www.astro.caltech.edu/~dperley/programs/autoastrometry.py}}, using the USNO B1.0 catalog as a reference \citep{usno}, with 0.32\arcsec~accuracy. 

\begin{figure*}[!ht]
\begin{center}
\begin{tabular}{c}
\includegraphics[scale=0.48]{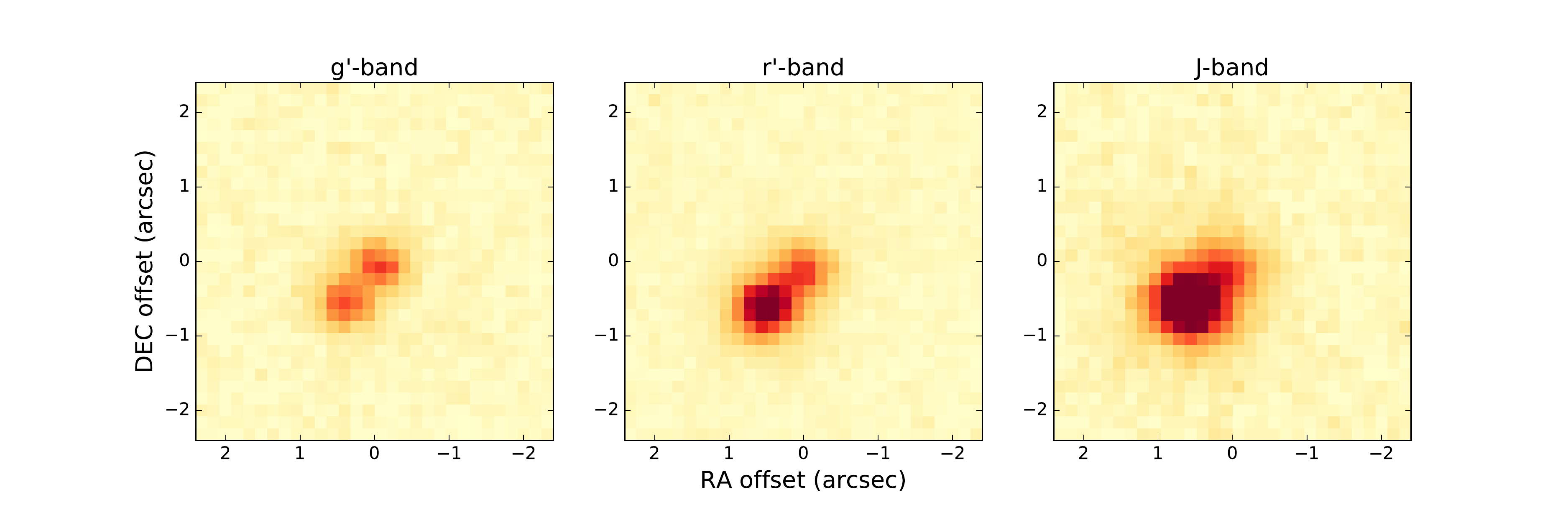}  \vspace{-0.3in} \\

 \hspace{0.2 in} \includegraphics[scale=0.65]{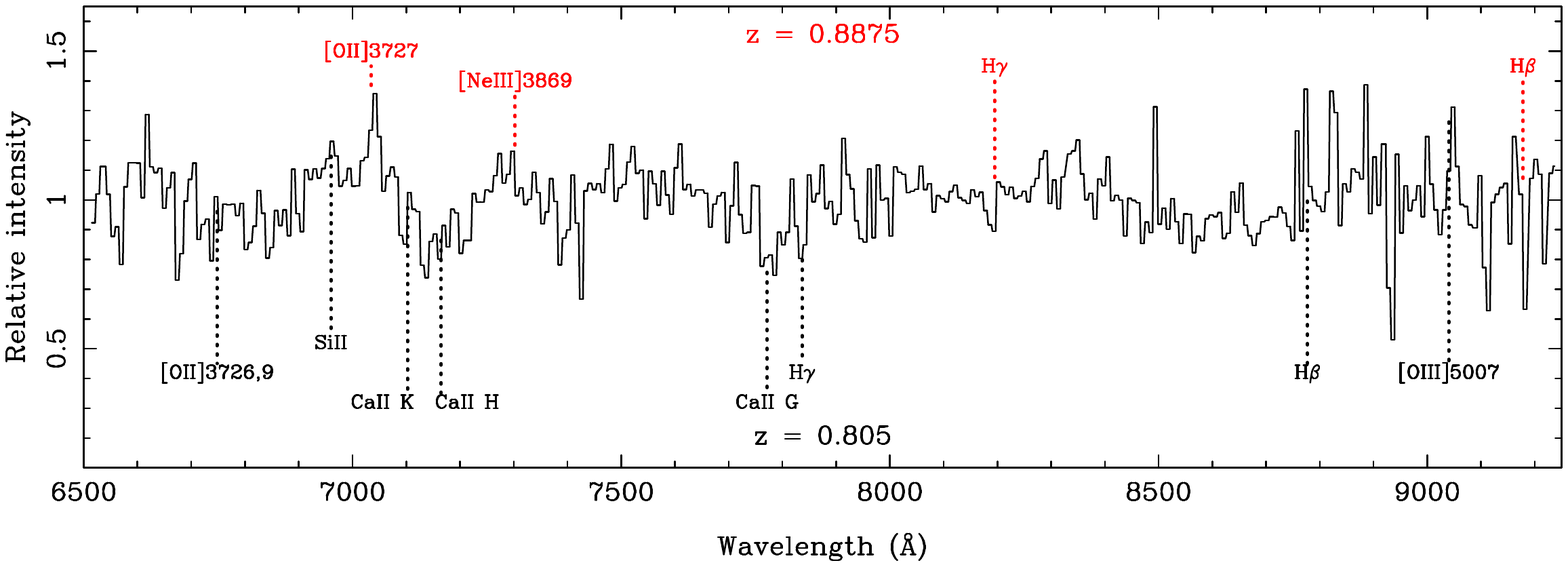}  \vspace{-3in}
\end{tabular}
\end{center}
\caption{ Upper: images of optical counterpart and its pair in $g^\prime$, $r^\prime$ (GMOS-South), and $J$ bands (Magellan).  Source G1, 
which is the counterpart to AT~J0642.9$-$5118, is in the image center, and G2 is to the lower left (south east). 
The intensity scaling (zero-point) is common between the images. Bottom:  Continuum-divided combined spectrum obtained of the pair of optical 
sources.  We show features potentially associated with source G1 in black and source G2 in red, with the corresponding redshifts quoted in the figure. 
Note that numerous artefacts due to imperfect sky-line subtraction are present. The instrumental FWHM of the spectrograph was 3\,\AA. 
 \label{fig:optical_image}  
    }
\end{figure*}

We identified a point-like counterpart to the radio source that is the north-west component of a close (0.6\arcsec~separation) double (Fig.~\ref{fig:optical_image}, left and 
middle panels). We term this source G1, and its south-eastern companion G2. As we did not observe a photometric standard field, we used the GMOS-South 
photometric equation defined online\footnote{\url{https://www.gemini.edu/sciops/instruments/gmos/calibration/photometric-stds}} to set the flux scale. We modeled the 
point-spread function using nearby stars and used this to model G1 and G2 as two point sources, finding a satisfactory fit to the observation.
For  G1, we obtained AB magnitudes of $g'=22.82\pm0.02$ and $r'=22.51\pm0.02$, and for G2 we obtained $g'=22.87\pm0.02$ and $r'=21.77\pm0.01$. At this position, 
the Galactic extinction is 0.208 magnitudes in the $g'$ band, and 0.144 magnitudes in the $r'$ band \citep{extinction}.

We also obtained imaging observations in the $J$ band with the FourStar instrument \citep{fstar} on the Magellan-Baade telescope at Las Campanas Observatory. 
The observations, conducted on MJD~57270 under photometric conditions with 0.65\arcsec~FWHM seeing, were split into 18 dithered exposures totalling 1153\,s. 
The data were reduced using the standard FourStar pipeline. We calibrated the photometry and astrometry of the image using 2MASS point sources \citep{tmass}, 
attaining 0.2\arcsec~astrometric accuracy. The resulting detections of G1 and G2 are shown in the right panel of Fig.~\ref{fig:optical_image}. Using the 
same technique as above, we measure AB magnitudes of $J=21.54\pm0.03$ for G1, and $J=20.24\pm0.01$ for G2. The Galactic extinction in the $J$ band is 
0.045 magnitudes \citep{extinction}. 

The point source catalog of the Widefield Infrared Survey Explorer \citep[WISE;][]{wise} contains a source, WISE~J064257.16-511817.8, which is coincident with G1 and is 
detected in the two shortest wavelength bands. Its (AB) magnitudes are $W1 = 17.9\pm0.1$ and $W2 =16.7\pm0.2$. 
Based on this color, the source is consistent with an active galactic nucleus \cite[AGN;][]{2012ApJ...753...30S}. The optical colors are also 
consistent with an AGN at moderate redshift, such that Ly$\alpha$ is blue-ward of our observations \citep{smolcic}. 

\subsection{Spectroscopy}

We also obtained optical spectra of G1 and G2 using GMOS at Gemini-South. We used a 1\arcsec~longslit oriented along the axis of G1 and G2 (position angle of 317$^{\circ}$). 
Our observations in the red part of the spectrum were conducted on MJD~57362 using the R400 grating with the GG455 order-blocking filter. Four 920\,s exposures 
were taken at a mean airmass of 1.2, with two centered on 8610\,\AA~and two centered on 8510\,\AA~to cover the gaps between CCDs. Our blue 
observations were conducted on MJD~57367 at a mean airmass of 1.1 using the B600 grating with no filter, and three 920\,s exposures (two centered on 5060\,\AA, and 
one centered on 4960\,\AA). We reduced the data using the standard GMOS pipeline, with a bias observation obtained on MJD~57363, and flat-field observations 
taken in between our science exposures. 

Unfortunately, intermittent high cirrus was present, precluding accurate spectrophotometric calibration and making sky emission lines 
difficult to subtract. We nonetheless used observations of a spectrophotometric calibrator on MJD~57562 (LTT~3218) to calibrate telluric absorption features. 
As the seeing on both nights was $\sim1$\arcsec~FWHM, as measured from acquisition images, we could not deconvolve the spectral traces of G1 and G2. Furthermore, 
some light from G1 was likely refracted out of the slit as GMOS does not contain an atmospheric dispersion corrector. Hence, following optimal extraction of the spectra, 
we only considered data taken using the R400 grating at wavelengths shorter than 9250\,\AA~and divided the data by a smooth polynomial fit to the continuum. 
The resulting spectrum, which contains numerous artifacts due to imperfect sky subtraction and is binned to 8\,\AA~resolution, is shown in Fig.~\ref{fig:optical_image}. 

We tentatively identify two redshifted systems in this spectrum: one at $z=0.805\pm0.001$ and one at $z=0.8875\pm0.001$. The first system is consistent with the spectra of 
early-type galaxies \citep{swire}, a hypothesis which is additionally evidenced by a possible spectral break in the continuum around 4000\,\AA. We thus interpret it as corresponding to 
G2. The redshift of the second system, presumably G1 (the radio source), is estimated primarily using the strong emission line at 7040\,\AA~and by assuming (based on the WISE  and blue optical colors) that it is an AGN. Identifying the 7040\,\AA~line with [OII]3272\,\AA~results in a clear prediction, specifically that strong emission lines (e.g., Ly$\alpha$, CIV, MgII) 
should be seen at shorter wavelengths. It appears to exhibit H$\gamma$ and H$\beta$ in absorption. The lack of these normally broad lines in emission, combined with 
the compact nature of its radio counterpart evidenced by the variability, is suggestive of a narrow-line radio galaxy \citep{o78}, or perhaps a radio-loud 
narrow-line Seyfert 1 \citep{nlsy}. More sensitive spectra with broader wavelength coverage would help in this classification, for example by searching for the  FeII emission features that distinguish the 
narrow-line Seyfert 1 class.

\section{Discussion}
\label{sec:discuss} 

\subsection{Limits on afterglows from the $\gamma$-ray pulse}
\label{sec:noafterglow}

Our observations can be used to search for afterglow emission associated with the potential $\gamma$-ray transient Swift J0644.5$-$5111.
In the classic fireball model \cite[][]{2000ApJ...537..191F}, the flux density of radio synchrotron emission is directly related to the input energy.
\cite{2016arXiv161103848M} calculated the flux density assuming the spectrum is not self absorbed and that the frequency of interest is below the peak  of the spectrum so that the flux density is still rising.  This is a reasonable assumption for our observations within $9$ and $10$~d after the FRB.  Assuming a distance $D=3.3$~Gpc consistent with the pulses extragalactic dispersion measure,  after time $T=10$~d at a frequency $\nu=5.5$~GHz, the flux density is
\begin{eqnarray}
S_\nu &&= 470~\mu{\rm Jy}  \left( \frac{\nu}{\rm 5.5~GHz} \right)^{1/3}  \left( \frac{E_\gamma}{10^{51.7}~{\rm erg}}\right)^{5/6}  \nonumber \\  
&&~~~\times \left( \frac{n}{0.1~{\rm cm}^{-3}}\right)^{1/2} \left( \frac{\epsilon_B}{10^{-2}} \right)^{1/3} \left( \frac{\epsilon_e}{0.1}\right)^{-2/3} f_e^{5/3}  \nonumber \\
&& ~~~\times  g(2.4) ^{-2/3}  \left(\frac{T}{\rm 10~d} \right)^{1/2} \left( \frac{D}{3~{\rm Gpc}}\right)^{-2},
\end{eqnarray}
where $E$ is the energy emitted in $\gamma$-rays, $n$ is the electron number-density of the shocked medium, $\epsilon_B$ is the magnetic energy fraction, $\epsilon_e$ is the nonthermal-electron energy density, $f_e$ is the nonthermal electron energy, and $g(s) =(s-2)/(s-1)$.
For fiducial assumptions for these parameters, we could have detected the source in both the $5.5$~GHz and $7.5$~GHz observations $10$~d after the 
explosion, with a significance of $12-20\sigma$.
This suggests that either the input luminosity is smaller than estimated in \cite{carinaswift},  that the environment surrounding the burst is unlike that of long $\gamma$-ray bursts or core-collapse supernova explosions, or that the $\gamma$-ray transient is unrelated to the FRB or spurious.

\subsection{The ATCA variable source}
\label{sec:variable_comp}

We interpret the variable radio source AT~J0642.9$-$5118 as  emission from compact components in a radio-loud AGN.
 This is evidenced by the persistent radio variability on timescales 
of days to months, the optical to mid-infrared colors of its host system, and its possible spectral identification. The lightcurve of the flare following FRB~131104, with the spectrum 
inverting when it brightens, is consistent with the classic picture of an expanding and cooling synchrotron bubble.
 Although the flare lightcurve is consistent with the radio 
afterglows of relativistic transients \citep[e.g.,][]{2000ApJ...537..191F}, the persistence and low-level variability of the radio source beyond the flare means that we 
have no evidence to favor a transient coinciding with a variable radio source, over simply a variable radio source. 

AT~J0642.9$-$5118 is nonetheless interesting.  This 
object has substantial differences from the variable radio source identified with FRB~150418 \citep{2016Natur.530..453K}. 
First, we clearly identify the flare of AT~J0642.9$-$5118 with the days immediately after the FRB, as we observe the flux density rise and the spectrum invert. 
Additionally, scintillation in the Milky Way interstellar medium is less likely  to cause the variability 
of AT~J0642.9$-$5118 because the source is at a relatively high galactic latitude ($b=-22^\circ$), and the scattering is expected to be in the weak regime 
\citep[e.g.,][]{w98} at 7.5~GHz.  It is possible that the source was magnified by an extreme scattering event 
\cite[][]{2016Sci...351..354B}, but that is improbable as only one in $\approx 2000$ compact sources are undergoing one at a given time.
Perhaps most importantly, AT~J0642.9$-$5118 has not re-brightened to within a factor of two of its flux densities as the peak of the flare, unlike the case for FRB~150418  
\cite[][]{2016ApJ...821L..22W,2016arXiv161009043J}. The flare of AT~J0642.9$-$5118 following FRB~131104 thus appears to be a transient 
occurrence within the scope of our monitoring of its flux density.  

Blind surveys for transients at lower frequencies find objects with such extreme variability (factor of three on few-month timescales) 
only very rarely \citep[e.g.,][]{mooley}. This is not surprising. 
Assuming constant brightness temperature, 
intrinsic AGN variability timescales  scale with frequency proportional to  $\approx\nu^{-1}$. Transient AGN flare events at higher frequencies are hence 
generally expected to be shorter in time, and are often also larger in modulation, than at lower frequencies \citep[e.g.,][]{hnt+08}. 
Scintillation timescales in the strong scintillation regime are expected to be more rapid. The post-FRB flare 
of AT~J0642.9$-$5118 is clearly most dramatic at the highest observing frequency.
 
 The temporal coincidence of  AT~J0642.9$-$5118 flare with FRB~131104 nonetheless motivates us to consider the possibility that it is associated 
 with the FRB. In this case, AGN activity would be implicated in FRB production. 
The potential redshift ($z=0.8875$) of AT~J0642.9$-$5118 is consistent with the extragalactic DM of the FRB \citep{dolag15}. 
While other source channels have been more strongly advocated for FRBs, it is not implausible that AGN could produce FRBs.
 Millisecond-duration radio pulses propagating through relativistic plasma in AGN jets may be immune to both absorption and scattering effects 
\citep{l08}, implying that FRBs originating close to launching regions could be observed from AGN viewed along the jets. 
Mechanisms \citep[e.g.,][]{romero16} have been 
proposed for the production of FRBs in AGN jets, analogous to the mechanisms for generating TeV photons. 

There are however reasons to disfavor an association between AT~J0642.9$-$5118 and the FRB.
The background transient and variable event rate at $7.5$~GHz (where the flare is the most prominent), and hence the false-alarm rate 
for the association, is poorly constrained. 
Even so,  the  FRB rate needs to be reconciled with the background rate \cite[][]{2016ApJ...824L...9V}.
Intrinsic AGN variability is likely to dominate the background slow-transient rate. 
A detailed 
analysis of the radio AGN population and its variability properties in comparison with the FRB rate would be required to assess how 
commonly a single object would be expected to emit an FRB, and what its signature could be. Further physical modeling of the conditions 
and orientations under which FRBs could escape AGN would help refine such an analysis. 
This analysis would be further aided by a large area survey for transient and variable sources at high frequencies, as well as dedicated follow-up 
observations of other FRBs, in particular to assess the frequency of short-duration flares in AGN. 
However, a substantially more constraining result would be the direct interferometric localization of a  population of FRBs  to flaring AGN.


\section{Conclusions}
\label{sec:conclude}

We present 25 epochs of centimetric imaging observations of the field of FRB~131104 with the Australia Telescope Compact Array spanning 2.5\,yr. 
No radio afterglow is coincident with the $\gamma$-ray event reported by \citet{carinaswift}. This tightly constrains the energetics the associated 
cataclysm, or suggests, as supported by probabilistic arguments we outline, that the $\gamma$-ray event is unrelated to the FRB or spurious.

We have identified an unusual flaring radio source temporally and spatially coincident with FRB~131104. This source, AT~J0642.9$-$5118, is not spatially 
coincident with the potential $\gamma$-ray transient Swift J0644.5$-$5111.
AT~J0642.9$-$5118 is consistent with compact emission components in an AGN, as identified by optical and infrared photometry and spectroscopy. 
The discovery of further, better-localized FRBs with either radio or $\gamma$-ray flares (or neither) will resolve the uncertainty  (or not) in the multiwavelength associations with 
the enigmatic fast radio burst population.

\acknowledgements
We thank M. Kasliwal for obtaining and reducing the J-band Magellan data presented in this paper, H. Vedantham, S. Kulkarni, 
K. Masui, R. Blandford, and S.~Johnston for useful discussions, and S. Ryder and the International Telescope Support Office at the Australian Astronomical Observatory for assistance in coordinating Gemini observations. We thank the group of M. Bailes at the Swinburne University of Technology for making available their real-time FRB detector, without 
which the rapid follow-up observations of FRB~131104 would not have been possible. We are also grateful for the prompt scheduling of our observations by CSIRO Astronomy and Space Science 
operations staff. 
The Australia Telescope Compact Array and Parkes radio telescope are part of the Australia Telescope which is funded by the Commonwealth of Australia for operation as a National Facility managed by the Commonwealth Science and Industrial Research Organization (CSIRO). 
This paper is partially based on observations obtained at the Gemini Observatory, which is operated by the Association of Universities for Research in Astronomy, Inc., under a cooperative agreement with the NSF on behalf of the Gemini partnership: the National Science Foundation (United States), the National Research Council (Canada), CONICYT (Chile), Ministerio de Ciencia, Tecnolog\'{i}a e Innovaci\'{o}n Productiva (Argentina), Minist\'{e}rio da Ci\^{e}ncia, Tecnologia e Inova\c{c}\~{a}o (Brazil), and, previously, the Department of Industry and Science (Australia).
This paper includes data gathered with the 6.5 meter Magellan Telescopes located at Las Campanas Observatory, Chile.

\facility{Parkes, ATCA, Gemini:South (GMOS), Magellan:Baade (FourStar)}


\end{document}